\batchmode
\makeatletter
\def\input@path{{D:/Data_IOP/ZrTe5/reply/ZrTe5_manuscipt_15/}}
\makeatother
\documentclass[twocolumn,prl,showpacs,superscriptaddress,longbibliography]{revtex4-2}
\usepackage[latin9]{inputenc}
\usepackage{color}
\usepackage{amstext}
\usepackage{graphicx}
\usepackage[pdftex,unicode=true,pdfusetitle,
 bookmarks=true,bookmarksnumbered=false,bookmarksopen=false,
 breaklinks=false,pdfborder={0 0 0},pdfborderstyle={},backref=false,colorlinks=true]
 {hyperref}


\begin{document}
\title{Temperature-Dependent Collective Excitations in a Three-Dimensional
Dirac System ZrTe$_{5}$}
\author{Zijian Lin}
\thanks{Equally contributed to this work.}
\affiliation{Beijing National Laboratory for Condensed Matter Physics and Institute
of Physics, Chinese Academy of Sciences, Beijing 100190, China}
\affiliation{School of Physical Sciences, University of Chinese Academy of Sciences,
Beijing 100049, China}
\author{Cuixiang Wang}
\thanks{Equally contributed to this work.}
\affiliation{Beijing National Laboratory for Condensed Matter Physics and Institute
of Physics, Chinese Academy of Sciences, Beijing 100190, China}
\affiliation{School of Physical Sciences, University of Chinese Academy of Sciences,
Beijing 100049, China}
\author{Daqiang Chen}
\affiliation{Beijing National Laboratory for Condensed Matter Physics and Institute
of Physics, Chinese Academy of Sciences, Beijing 100190, China}
\affiliation{School of Physical Sciences, University of Chinese Academy of Sciences,
Beijing 100049, China}
\author{Sheng Meng}
\affiliation{Beijing National Laboratory for Condensed Matter Physics and Institute
of Physics, Chinese Academy of Sciences, Beijing 100190, China}
\affiliation{School of Physical Sciences, University of Chinese Academy of Sciences,
Beijing 100049, China}
\affiliation{Songshan Lake Materials Laboratory, Dongguan, Guangdong 523808, China}
\author{Youguo Shi}
\affiliation{Beijing National Laboratory for Condensed Matter Physics and Institute
of Physics, Chinese Academy of Sciences, Beijing 100190, China}
\affiliation{Songshan Lake Materials Laboratory, Dongguan, Guangdong 523808, China}
\author{Jiandong Guo}
\email{jdguo@iphy.ac.cn}

\affiliation{Beijing National Laboratory for Condensed Matter Physics and Institute
of Physics, Chinese Academy of Sciences, Beijing 100190, China}
\affiliation{School of Physical Sciences, University of Chinese Academy of Sciences,
Beijing 100049, China}
\author{Xuetao Zhu}
\email{xtzhu@iphy.ac.cn}

\affiliation{Beijing National Laboratory for Condensed Matter Physics and Institute
of Physics, Chinese Academy of Sciences, Beijing 100190, China}
\affiliation{School of Physical Sciences, University of Chinese Academy of Sciences,
Beijing 100049, China}
\begin{abstract}
Zirconium pentatelluride (ZrTe$_{5}$), a system with a Dirac linear
band across the Fermi level and anomalous transport features, has
attracted considerable research interest for it is predicted to be
located at the boundary between strong and weak topological insulators
separated by a topological semimetal phase. However, the experimental
verification of the topological phase transition and the topological
ground state in ZrTe$_{5}$ is full of controversies, mostly due to
the difficulty of precisely capturing the small gap evolution with
single-particle band structure measurements. Alternatively, the collective
excitations of electric charges, known as plasmons, in Dirac systems
exhibiting unique behavior, can well reflect the topological nature
of the band structure. Here, using reflective high-resolution electron
energy loss spectroscopy (HREELS), we investigate the temperature-dependent
collective excitations of ZrTe$_{5}$, and discover that the plasmon
energy in ZrTe$_{5}$ is proportional to the $1/3$ power of the carrier
density $n$, which is a unique feature of plasmons in three-dimensional
Dirac systems or hyperbolic topological insulators. Based on this
conclusion, the origin of the resistivity anomaly of ZrTe$_{5}$ can
be attributed to the temperature-dependent chemical potential shift
in extrinsic Dirac semimetals.
\end{abstract}
\maketitle
\emph{Introduction.}---Topological semimetals and insulators have
attracted extensive attention due to their potential to achieve various
novel quantum states \citep{sarmaMajoranaZeroModes2015,lutchynMajoranaZeroModes2018}.
In principle, just as their names suggest, topological semimetals
are characterized by linear semimetallic bulk band crossings \citep{burkovTopologicalSemimetals2016},
while topological insulators have parabolic bulk bands with insulating
gaps \citep{hasanColloquiumTopologicalInsulators2010,xiaObservationLargegapTopologicalinsulator2009}.
Typically, these two kinds of band features can be well distinguished
by single-particle imaging techniques such as scanning tunneling spectroscopy
(STS) and angle-resolved photoemission spectroscopy (ARPES) \citep{lvExperimentalPerspectiveThreedimensional2021,lvAngleresolvedPhotoemissionSpectroscopy2019}.
From a theoretical perspective, topologically distinct insulating
phases, e.g., strong and weak topological insulators, are separated
by a topological semimetal phase \citep{murakamiPhaseTransitionQuantum2007a}.
The transition between these phases is defined as a topological phase
transition \citep{murakamiUniversalPhaseDiagrams2008}. Searching
for the topological phase transition in a real material system is
one of the most appealing topics in the fast-developing field of topological
physics \citep{castelvecchiStrangeTopologyThat2017}.
\begin{figure}
\includegraphics[width=8.6cm]{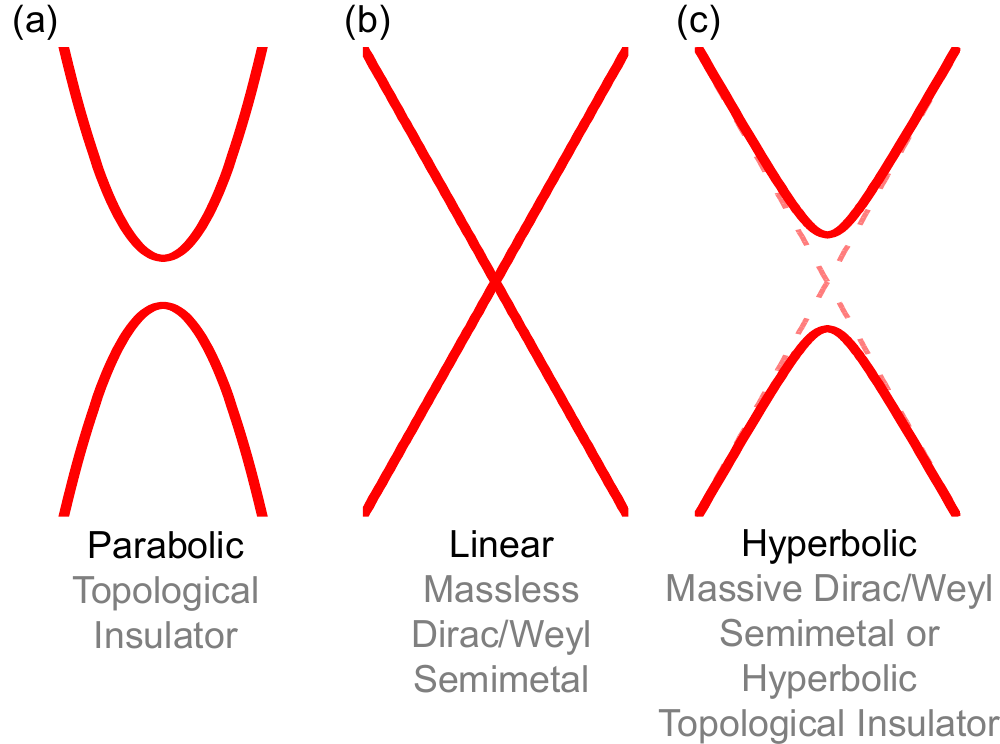}

\caption{\label{fig:Classification-of-bulk}Classification of bulk bands in
topological systems. (a) Conventional topological insulator with parabolic
bands. (b) Massless Dirac/Weyl semimetal with linear bands. (c) Massive
Dirac/Weyl semimetal or hyperbolic topological insulator with hyperbolic
bands (distinguished by the absence or presence of topological surface
states, respectively). The dashed line represents the asymptote of
the hyperbola.}
\end{figure}

ZrTe$_{5}$ is such a system that is theoretically predicted to be
at the boundary of a strong and weak topological insulator phase transition
\citep{wengTransitionMetalPentatellurideZrTe52014}, but experimentally
the verification of the topological phase transition is still controversial.
On one hand, the anomalies of resistivity \citep{okadaGiantResistivityAnomaly1980,stillwellEffectElasticTension1989,jonesThermoelectricPowerHfTe51982},
thermopower \citep{okadaGiantResistivityAnomaly1980,rubinsteinHfTe5ZrTe5Possible1999},
and Hall coefficient \citep{okadaGiantResistivityAnomaly1980,chiLifshitzTransitionMediated2017}
in ZrTe$_{5}$ seem to imply a temperature-induced topological phase
transition \citep{Xu2018}. However, on the other hand, there is currently
no consensus on the experimental band gap measurements. For example,
ARPES measurements \citep{liChiralMagneticEffect2016,manzoniEvidenceStrongTopological2016,shenSpectroscopicEvidenceGapless2017,Zhang2017,zhangObservationControlWeak2021}
report inconsistent gap values ranging from 0 to 100 meV at low temperatures
due to the difference in the data processing methods to deal with
the in-band broadening. And STS experiments face challenges in distinguishing
the V-shaped semimetal state \citep{manzoniEvidenceStrongTopological2016,shenSpectroscopicEvidenceGapless2017}
from a band gap \citep{liExperimentalObservationTopological2016,salzmannNatureNativeAtomic2020},
owing to the zero density of states at the Fermi level in a Dirac
semimetal phase.

In topological systems, the bulk bands can manifest in three different
shapes (illustrated in Fig. \ref{fig:Classification-of-bulk}): a
parabolic band with a gap (conventional topological insulator \citep{xiaObservationLargegapTopologicalinsulator2009,hasanColloquiumTopologicalInsulators2010}),
a gapless linear band (Dirac or Weyl semimetal \citep{dassarmaCollectiveModesMassless2009}),
or a hyperbolic band (massive Dirac semimetal \citep{Sachdeva2015}
or hyperbolic topological insulator). The hyperbolic band emerges
after continuously introducing a gap to a gapless linear band, and
the asymptotes of the hyperbolic curve trace back to the original
linear band. In principle, a strict linear band with a gap is theoretically
implausible. Therefore, even if the small gap size could be precisely
measured, it is still hard to determine if ZrTe$_{5}$ is a conventional
parabolic topological insulator, a massive Dirac semimetal, or a hyperbolic
topological insulator. The ground topological state of ZrTe$_{5}$
and the existence of its topological phase transition are still elusive.

In this context, attempting to determine the topological ground state
of ZrTe$_{5}$ from the perspective of gap size has to deal with various
complexities. Alternatively, investigating topological properties
from the perspective of collective excitations has emerged as a potential
approach, bypassing the controversies of direct single-particle gap
measurements, as collective excitations can uniquely reflect the features
of topological band structures \citep{dassarmaCollectiveModesMassless2009,stauberPlasmonicsDiracSystems2014}.
For example, a previous theoretical work suggests that Dirac plasmons,
which are collective excitations of linear Dirac electrons, have a
unique carrier density-dependent plasmon feature that is completely
different from that of the topological insulator with parabolic bulk
electrons \citep{Hofmann2015}. The advantage of this approach is
that it allows for a \emph{qualitative} assessment of the topological
ground state solely based on the shape of the energy band near the
Fermi level obtained from the plasmon behavior, without requiring
precise measurements of the gap size.

In this study, we employ reflective high-resolution electron energy
loss spectroscopy (HREELS) to investigate the temperature-dependent
collective excitations of ZrTe$_{5}$ to determine its topological
ground state. We observe the existence of plasmon-phonon coupling
(PPC) in ZrTe$_{5}$ and extract the temperature-dependent plasmon
energy via a phenomenological model. We discover that the plasmon
energy in ZrTe$_{5}$ exhibits a distinct behavior, proportional to
$n^{1/3}$ (where $n$ is the carrier density), which is a characteristic
feature of plasmons in three-dimensional (3D) Dirac systems with linear
or hyperbolic Dirac bands \footnote{The possibility of a hyperbolic Dirac band with non-zero mass cannot
be excluded, which will be discussed in the SM \citep{SI}.}. This rules out the possibility that ZrTe$_{5}$ is a conventional
topological insulator with parabolic bulk bands. Besides, we discuss
the potentiality of it being a topological insulator with hyperbolic
bulk bands. Based on these findings, we attribute the origin of the
resistivity anomaly in ZrTe$_{5}$ to an extrinsic-intrinsic crossover
of the Dirac electron behavior, without necessitating the additional
assumption of a topological phase transition.

\emph{Method.---}The single crystals of ZrTe$_{5}$ were grown using
the chemical vapor transport (CVT) method. ZrTe$_{5}$ single crystals
were cleaved in an ultra-high vacuum and the crystallographic orientations
were checked by low-energy electron diffraction. Then the collective
excitations were measured $in$ $situ$ in an HREELS system with a
reflected scattering geometry \citep{zhuHighResolutionElectron2015}.
The incident electron beam energy $E_{\text{i}}=110$ eV and the incident
angle $\theta=65^{\circ}$ were used with a typical energy resolution
of 2 meV. The data were collected with the sample temperature varying
from 35 to 300 K at $\overline{\Gamma}$ point ($q_{\parallel}=0$)
in the surface Brillouin zone. The band structure of ${\rm ZrTe_{5}}$
was calculated by the first-principles density functional theory (DFT)
as implemented in the \textsc{Quantum ESPRESSO} package \citep{giannozziQUANTUMESPRESSOModular2009}.
The plasmons were obtained from the calculations of the dielectric
functions as well as the energy-loss functions within the random phase
approximation. The details about the experimental methods, sample
characterizations, and plasmon calculations are described in the Supplemental
Material (SM) \citep{SI}.

\begin{figure}
\noindent \begin{raggedright}
\includegraphics[width=8.6cm]{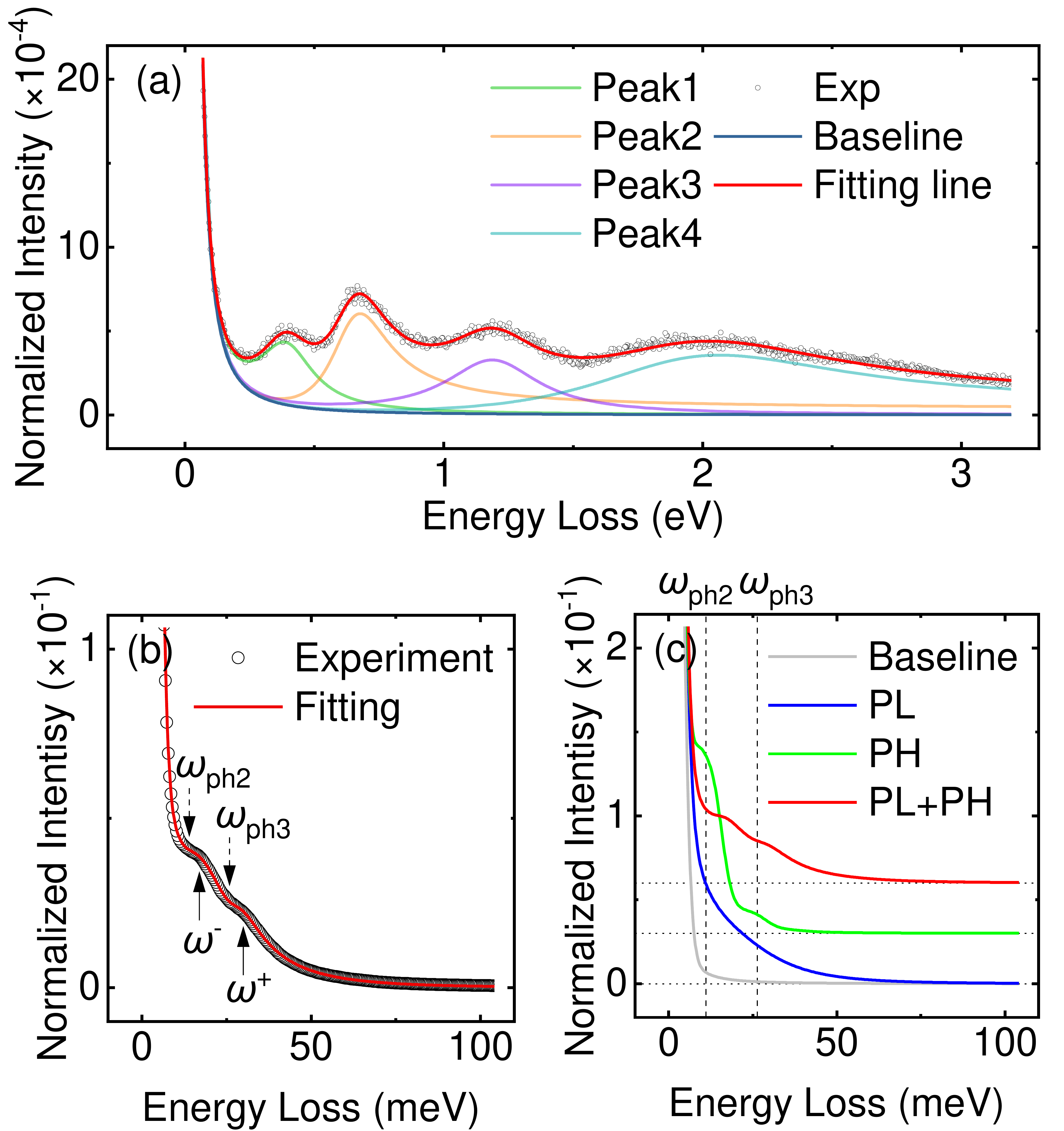}
\par\end{raggedright}
\caption{\label{1}HREELS results of ZrTe$_{5}$ at 300 K. (a) High energy
excitations. The black circles are the energy distribution curve (EDC)
from the HREELS experiment and the red line is the fitting result
using a baseline (blue), two Lorentz peaks (green and purple), and
two Fano peaks (orange and cyan). (b) Low energy excitations. The
black circles are the EDC from the HREELS experiment and the red line
is the fitting result using the Drude-Lorentz model. The features
of plasmon peak splitting $\omega^{\pm}\approx16$ and $28$ meV induced
by PPC and the characteristic dip caused by phonons $\omega_{\text{ph}}\approx10.5$
and $25.8$ meV are represented by upward solid arrows and downward
dashed arrows, respectively. (c) Features of PPC in the fitting EDC.
The gray line is the baseline due to the elastic scattering without
any sample information. The blue and the green lines are the fitting
results with only one plasmon (PL) or with only three phonons (PH),
respectively. The red line {[}same as the red line in panel (b){]}
is the fitting with one plasmon and three phonons (PL+PH). Horizontal
dotted lines show the offset between the different fitting results.
The vertical dashed lines, $\omega_{\text{ph2}}$ and $\omega_{\text{ph3}}$,
represent the energies of the phonons in the fitting.}
\end{figure}
\emph{Plasmon-phonon coupling.}---Figure \ref{1} displays the energy
distribution curves (EDCs) of the HREELS spectra in the high loss
energy region (\textless{} 3 eV) and low loss energy region (\textless{}
100 meV), respectively. In the high energy region, four excitations
around 0.4, 0.6, 1.2, and 1.9 eV were observed {[}Fig. \ref{1}(a){]}.
With the analysis of the calculated density of states and the dielectric
functions (see details in the SM \citep{SI}), the three excitations
around 0.4, 0.6, and 1.2 eV are assigned to interband transitions,
and the excitation around 2.0 eV is identified as an interband plasmon.
The temperature dependence of these excitations was not found to be
related to the Dirac electrons. Instead, they are demonstrated to
be originating from thermally induced lattice expansion (see details
in the SM \citep{SI}). Therefore, the focus of this work is on the
results in the low energy (\textless{} 100 meV) region.

Figure \ref{1}(b) shows the low energy HREELS results in ZrTe$_{5}$
at room temperature (300 K). Two subtle features at $\hbar\omega=$
16 and 28 meV are detected. Previous works \citep{chenOpticalSpectroscopyStudy2015,Xu2018,martinoTwoDimensionalConicalDispersion2019}
reported the existence of three IR-active phonons with energies of
about 5, 11, and 23 meV and a plasmon with an energy of about 30 meV.
If these excitations are independent of each other, they should have
appeared as peaks centered at the phonon or plasmon energies in HREELS
spectra. However, the two characteristic peaks in ZrTe$_{5}$ are
inconsistent with the energies of the independent phonons and the
plasmon in previous IR measurements \citep{chenOpticalSpectroscopyStudy2015,Xu2018,martinoTwoDimensionalConicalDispersion2019}.
This inconsistency is caused by the PPC via a macroscopic electric
field \citep{matzConductionBandSurfacePlasmons1981}, which has been
well-studied in graphene \citep{Liu2010,Koch2010,kochRobustPhononPlasmonCoupling2016}
and doped semiconductors \citep{kaplanInfraredAbsorptionCoupled1967,mcmahonInfraredReflectivityDoped1969,olsonLongitudinalOpticalPhononPlasmonCouplingGaAs1969,kukharskiiPlasmonphononCouplingGaAs1973,kozawaRamanScatteringPhonon1994}.
In an HREELS spectrum, the PPC is typically observed as split plasmon
signals marked by $\omega^{\pm}$, where a plasmon contributes an
envelope and a phonon creates a dip on this envelope \citep{matzConductionBandSurfacePlasmons1981,Liu2010,Koch2010,kochRobustPhononPlasmonCoupling2016}.

The measured HREELS spectra were fitted to demonstrate the splitting
mechanism induced by the PPC. In the fitting process, we first consider
the dielectric function described by the Drude-Lorentz model, which
can describe the PPC via a macroscopic electric field, and subsequently
simulate the HREELS spectra from the dielectric function using a well-established
numerical method \citep{Lambin1990}. The dielectric function is expressed
as follows:

\[
\epsilon(\omega)=\epsilon_{\infty}-\frac{\Omega_{\text{\emph{p}}}^{2}}{\omega^{2}+{\rm i}\Gamma_{\text{\emph{p}}}\omega}+\sum_{k=1}^{3}\frac{Q_{k}\omega_{\text{ph}k}^{2}}{\omega_{\text{ph}k}^{2}-\omega^{2}-{\rm i}\Gamma_{\text{ph}k}\omega}.
\]
The first term $\epsilon_{\infty}$ on the right-hand side is the
high-frequency dielectric constant. The second term is the Drude term
describing plasmons, and the third term is the Lorentz term describing
the IR-active phonon modes, where $\Omega_{\text{\emph{p}}}$, $\Gamma_{\text{\emph{p}}}$,
\emph{Q}, $\omega_{\text{ph}}$, and $\Gamma_{\text{ph}}$ represent
the plasmon frequency, the carrier damping, the phonon strength, the
phonon frequency, and the phonon damping, respectively. The fitting
line {[}the red line in Fig. \ref{1}b){]} agrees well with the EDCs
measured by HREELS, and the fitting results are given as the screened
plasmon energy $\hbar\omega_{\mathcal{\text{\emph{p}}}}=\hbar\Omega_{\text{\emph{p}}}/\sqrt{\epsilon_{\infty}}=31.5$
meV, the phonon energies $\hbar\omega_{\text{ph1}}=5.5$ meV, $\hbar\omega_{\text{ph2}}=10.5$
meV, and $\hbar\omega_{\text{ph3}}=25.8$ meV (see other parameters
in the SM \citep{SI}). The energies of phonons and the plasmon obtained
from this fitting show little discrepancy to the IR experimental results,
and meanwhile the fitting successfully reproduces the splitting features
marked by $\omega^{\pm}$ {[}Fig. \ref{1}(b){]}.

In order to establish an intuitive understanding of the splitting
features induced by the PPC, we also performed fittings with only
a plasmon considered or only three phonons considered. As shown in
Fig. \ref{1}(c), the gray line represents the background from the
elastic scattering without sample information. \textit{Case 1}: Considering
the Drude component from an independent plasmon only, the spectral
weight is added in the EDC as a swelling centered at the energy $\hbar\omega_{\text{\emph{p}}}=31.5$
meV (the blue line), which is so close to the zero loss peak that
it cannot manifest as a distinguishable peak. \textit{Case 2}: When
considering only three Lorentz components from phonons (the green
line), each phonon contributes a peak (or a shoulder) in the EDC,
except for the one with the energy $\hbar\omega_{\text{ph1}}=5.5$
meV merging into the zero loss peak. In summary, when plasmons and
phonons exist independently, they appear as peaks centered at their
respective energies. Both of these two cases exhibit significant discrepancies
compared to our experimental results. In contrast, when all the phonons
and the plasmon are included, the curve shape is significantly altered,
demonstrating the PPC spitting feature, as shown by the red lines
in Fig. \ref{1}(b) and (c). Specifically, the Drude component (the
plasmon) contributes an envelope centered at $\hbar\omega_{\text{\emph{p}}}=31.5$
meV, and the phonons provide dips on the envelope, as if the phonons
split the plasmon.

\begin{figure}
\noindent \begin{raggedright}
\includegraphics[width=8.6cm]{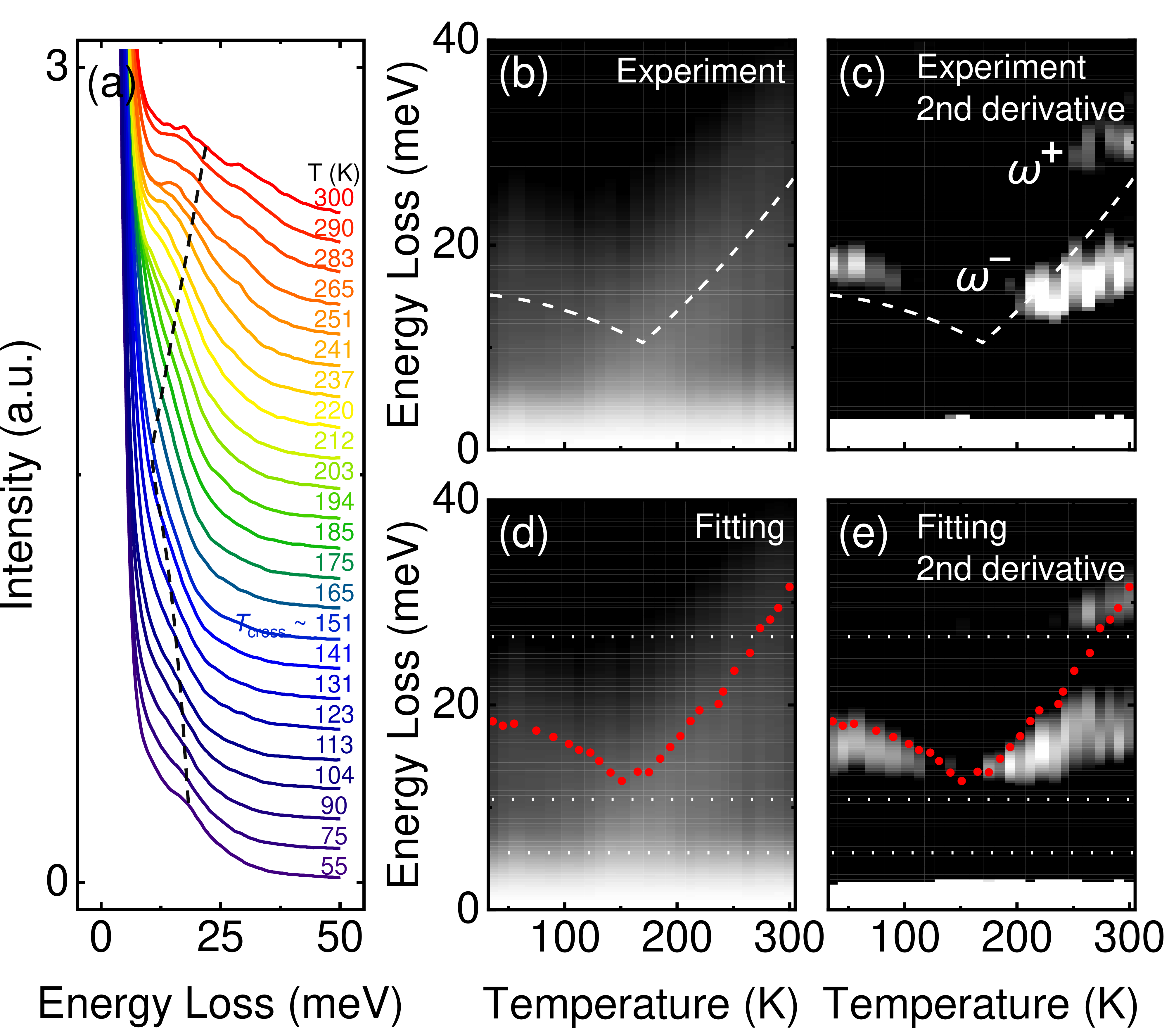}
\par\end{raggedright}
\caption{\label{2}Temperature dependence of the plasmon in ZrTe$_{5}$. (a)
Stack of EDCs at varying temperatures from 55 to 300 K. The dashed
line is the guide to the eyes of the envelope contributed by the plasmon
{[}same in panels (b) and (c){]}. (b) Color mapping of panel (a).
The color bar is linear {[}same in panels (c)-(e){]}. (c) Second derivative
mapping of panel (b). (d) and (e) Fitting results corresponding to
panels (b) and (c). The red circles are the fitted screened plasmon
energy $\hbar\omega_{\text{p}}$, and the white dotted line represents
the three phonon energies $\hbar\omega_{\text{ph1,2,3}}$, which are
determined through fitting at 300 K, and remain constant in the fitting
at other temperatures.}
\end{figure}
\emph{Temperature dependence of the plasmon.}---Based on the above
analysis, we can examine the temperature dependence of the plasmon
using the method of tracking the envelope center and fitting with
the Drude-Lorentz model. Figure \ref{2} presents the temperature-dependent
low-energy EDCs. Qualitatively, the central energy of the envelope
contributed by the plasmon declines with the decreasing temperature,
and then slightly rises below about 150 K {[}dashed lines in Fig.
\ref{2}(a) and (b){]}. The peak temperature for the plasmon energy
coincides with the peak temperature for the carrier density measured
in a sample grown by the CVT method \citep{martinoTwoDimensionalConicalDispersion2019},
establishing a resistivity peak temperature $T_{\text{cross}}\sim145$
K (details in the SM \citep{SI}). The color mapping of the second
derivative {[}Fig. \ref{2}(c){]} displays fine structures of the
plasmon splitting. The two modes above $\sim$ 250 K correspond to
the $\omega^{+}$ and $\omega^{-}$ in Fig. \ref{1}(a). With the
plasmon energy $\hbar\omega_{\text{\emph{p}}}$ decreasing upon cooling,
the plasmon moves further away from the phonon with energy $\hbar\omega_{\text{ph3}}$
and the PPC weakens, resulting in the less discernible $\omega^{+}$
mode below $\sim$ 250 K. Additionally, the plasmon should have been
split again by the phonon with energy $\hbar\omega_{\text{ph2}}$
when it approaches this phonon. The splitting mode with lower energy
is so close to the zero loss peak that it merges into it and therefore
cannot be resolved.

To further quantitatively determine the temperature dependence of
the plasmon, the aforementioned Drude-Lorentz fitting was conducted
while keeping the phonon energies unchanged (more fitting results
in the SM \citep{SI}). The fitting results {[}Fig. \ref{2}(d) and
(e){]} can unambiguously reproduce all the experimental features,
including the non-monotonic trend of the plasmon envelope {[}Fig.
\ref{2}(b) and (d){]} and the PPC induced-splitting {[}Fig. \ref{2}(c)
and (e){]}. The screened plasmon energy $\hbar\omega_{\text{\emph{p}}}$
as a function of temperature is now extracted with high precision.

\begin{figure}
\noindent \begin{raggedright}
\includegraphics[width=8.6cm]{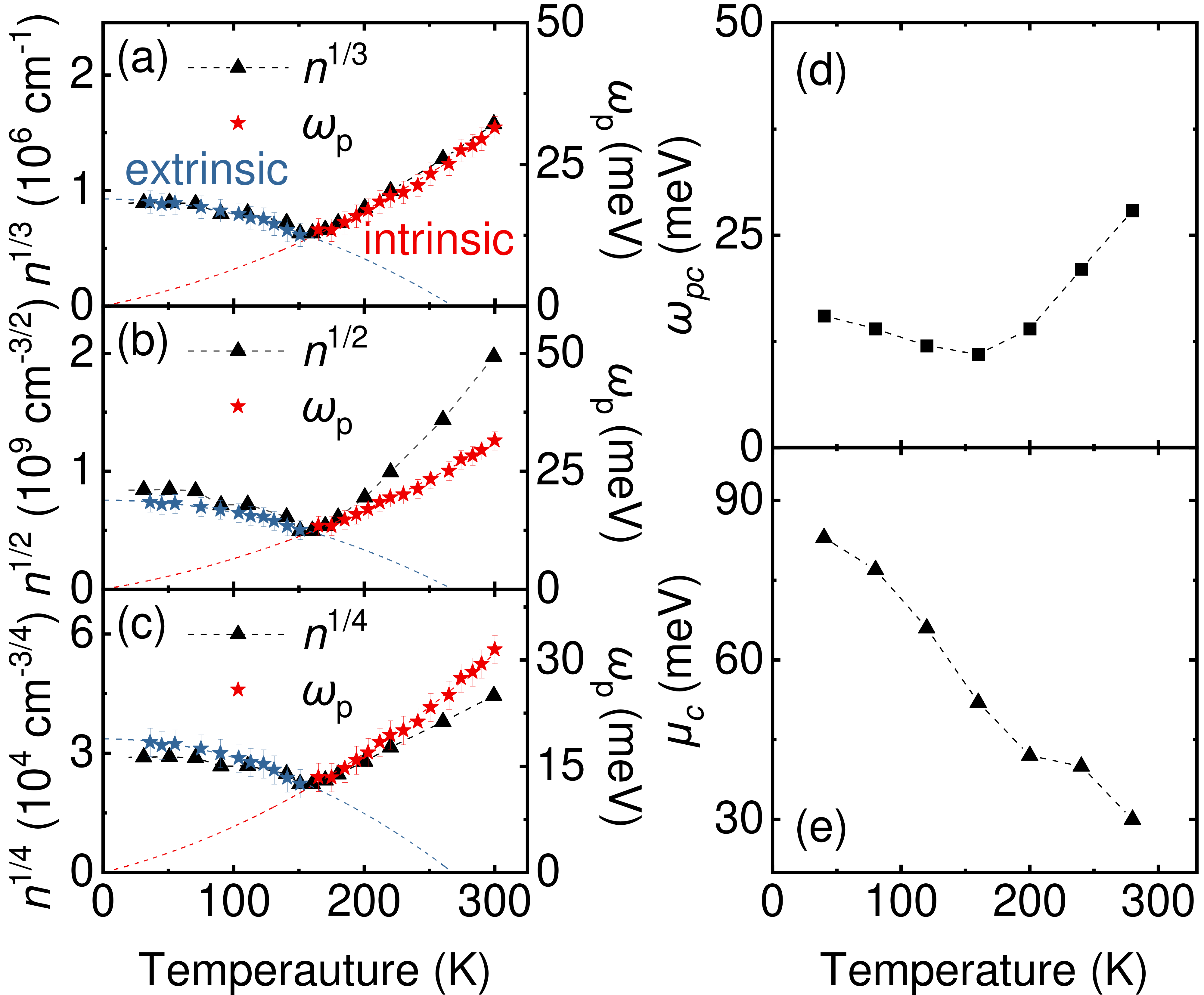}
\par\end{raggedright}
\caption{\label{3}Verification of the plasmon behavior in a 3D Dirac electron
system. (a)-(c) Comparison between the plasmon energy and the carrier
density in ZrTe$_{5}$. The stars (including the blue and red) represent
the screened plasmon energy $\omega_{p}$, and the triangles represent
the power of the carrier density (a) $n^{1/3}$, (b) $n^{1/2}$ and
(c) $n^{1/4}$, respectively. The data of $n$ are extracted from
Ref. \citep{martinoTwoDimensionalConicalDispersion2019}. The blue
and red dashed lines are the fitting results using theoretical models
in extrinsic and intrinsic Dirac systems \citep{Hofmann2015} of the
plasmon energy (details in the SM \citep{SI}), respectively. (d)
Calculated plasmon energy $\omega_{\text{pc}}$ at different temperatures.
(e) Temperature-dependent chemical potential $\mu_{c}$ used in the
calculation.}
\end{figure}

\emph{Plasmon in the 3D Dirac electron system.}---The above quantitative
temperature dependence can be utilized to clarify the topological
properties of ZrTe$_{5}$. In a 3D parabolic electron system {[}the
case in Fig. \ref{fig:Classification-of-bulk}(a){]}, the plasmon
energy is proportional to the carrier density to the power of 1/2,
i.e. $\hbar\omega_{\text{\emph{p}}}\sim n^{1/2}$ \citep{Bohm1953}.
However, in a massless Dirac electron system {[}the case in Fig. \ref{fig:Classification-of-bulk}(b){]},
the plasmon energy follows $\hbar\omega_{\text{\emph{p}}}\sim\sqrt{n}/n^{1/2d}$
\citep{Hofmann2015}, where $d$ is the dimension, yielding $\hbar\omega_{\text{\emph{p}}}\sim n^{1/3}$
for $d=3$ and $\hbar\omega_{\text{\emph{p}}}\sim n^{1/4}$ for $d=2$.
The comparison between the plasmon energy and the carrier density
is shown in Fig. \ref{3} (the carrier density data from Ref. \citep{martinoTwoDimensionalConicalDispersion2019}).
The plasmon energy is almost entirely described by $n^{1/3}$ {[}Fig.
\ref{3}(a){]} for 3D Dirac systems instead of $n^{1/2}$ {[}Fig.
\ref{3}(b){]} for 3D parabolic systems or $n^{1/4}$ for 2D Dirac
systems {[}Fig. \ref{3}(c){]}. In massive 3D Dirac systems {[}the
case in Fig. \ref{fig:Classification-of-bulk}(c){]}, $\hbar\omega_{\text{\emph{p}}}$
will deviate very slightly from $n^{1/3}$. Our results display a
very subtle trend, with the mass gradually increasing as the temperature
rises (see details in the SM \citep{SI}). This provides solid evidence
that the plasmon observed in ZrTe$_{5}$ is a collective excitation
composed of the 3D Dirac electrons. Thus the ground topological states
of ZrTe$_{5}$ could be either a 3D Dirac semimetal (including massless
or massive) or a hyperbolic topological insulator.

Our measurements rule out the possibility that ZrTe$_{5}$ is a conventional
topological insulator with parabolic bulk bands. Unfortunately, it
is still hard to distinguish between a massive Dirac semimetal and
a hyperbolic topological insulator, the latter of which has yet to
be explored either theoretically or experimentally.

The behavior that the plasmon energy and the carrier density first
decrease and then increase upon warming is an inherent property of
the 3D Dirac system. The extrinsic 3D Dirac system \citep{Hofmann2015},
where the chemical potential $\mu(T=0)$ locates above the Dirac point
at zero temperature. In an intrinsic Dirac system, where the zero-temperature
chemical potential $\mu(T=0)$ locates at the Dirac point, the carrier
density and the plasmon energy monotonically increase with the temperature
increasing following $\hbar\omega_{\text{\emph{p}}}\sim T/\text{\ensuremath{\sqrt{\ln(1/T)}}}$
due to the thermal excitation and ultraviolet renormalization \citep{Hofmann2015},
which is called the intrinsic effect of a Dirac system. However, in
an extrinsic Dirac system, the zero-temperature chemical potential,
$\mu(T=0)$ locates away from the Dirac point, offering finite carrier
density $n$ and plasmon energy $\hbar\omega_{\text{\emph{p}}}$.
As temperature increases, thermal excitations of carriers pull $\mu(T)$
towards the valance band in the condition of the particle number conservation,
and then the plasmon energy monotonically decreases following $\hbar\omega_{\text{\emph{p}}}\sim\mu(T)$
\citep{Hofmann2015}, which is called the extrinsic effect of the
Dirac system. Upon reaching sufficiently high temperatures, where
$\mu(T)$ is brought close to the valence band, the system undergoes
an extrinsic-intrinsic crossover around the cross temperature $T_{\text{cross}}$,
and eventually displays intrinsic effects at high temperatures. Overall,
an extrinsic Dirac system is dominated by the extrinsic and intrinsic
effects below and above $T_{\text{cross}}$, respectively. This phenomenon
has been previously observed in 3D Dirac systems Na$_{3}$Bi and Cd$_{3}$As$_{2}$
\citep{Jenkins2016}. Our first-principles calculations (see details
in the SM \citep{SI}) of ZrTe$_{5}$ have successfully reproduced
the non-monotonic dependence of the plasmon energy on temperature
{[}Fig. \ref{3}(d){]}, with $\mu(T)$ monotonically approaching the
valence band {[}Fig. \ref{3}(e){]}. When the mobility change is minimal
\footnote{In Ref. \citep{martinoTwoDimensionalConicalDispersion2019}, the carrier
type changes from electrons to holes across 135 K, since the chemical
potential shifts towards the pocket with smaller density of states
as the temperature increases. This can partially affect the resistance,
but has no impact on the total carrier density $n$.}, the resistivity follows $\rho\sim n^{-1}\sim\omega_{p}^{-3}$ in
a 3D Dirac system. Therefore, the resistivity anomaly {[}largest resistance
corresponding to the lowest $\hbar\omega_{p}$ at $T_{\text{cross}}${]}
can be primarily attributed to the shift in chemical potential and
the enhancement of thermal excitations, without the requirement of
a topological phase transition \citep{Xu2018}.

In the extrinsic region, the plasmon and the chemical potential follow
the relation $\omega_{p}(T)=\omega_{p}^{0}[1-\frac{\pi^{2}}{6}(\frac{T}{T_{\text{\emph{F}}}})^{2}]+O(T^{4}/T_{\text{\emph{F}}}^{4})$
and $\mu(T)=E_{F}-\frac{1}{3}\pi^{2}E_{\text{\emph{F}}}^{-1}T^{2}$
\citep{Hofmann2015}. Here, $\omega_{\text{\emph{p}}}^{0}$ represents
the plasmon frequency at zero temperature, and $T_{\text{\emph{F}}}$
is the Fermi temperature. By fitting the data in the extrinsic region,
we obtained $T_{\text{\emph{F}}}=343$ K and $\mu(T=0)=34.3$ meV.
Meanwhile, we also noticed that the linear optical conductivity is
truncated at about 70 meV near zero temperature \citep{martinoTwoDimensionalConicalDispersion2019},
which is consistent with $2\mu(T=0)$.

\emph{Discussion.}---The extrinsic-intrinsic crossover mechanism
can also be applied to samples grown by the self-flux method \citep{chenOpticalSpectroscopyStudy2015,martinoTwoDimensionalConicalDispersion2019},
resulting in a smaller $\mu(T=0)$ and a lower $T_{\text{cross}}$.
This is consistent with the smaller truncation energy in the optical
conductivity (about 28 meV \citep{martinoTwoDimensionalConicalDispersion2019})
and the lower $T_{\text{cross}}=88$ K \citep{chenOpticalSpectroscopyStudy2015,martinoTwoDimensionalConicalDispersion2019}
compared to the samples grown by the CVT method. Furthermore, these
differences between samples can be attributed to the influence of
the Te defects \citep{salzmannNatureNativeAtomic2020} on the value
of the chemical potential $\mu(T=0)$. In general, the dependence
of $T_{\text{cross}}$ on samples highlights the significant influence
of the chemical potential $\mu(T=0)$ in ZrTe$_{5}$, thereby consolidating
the self-consistency of the extrinsic-intrinsic crossover mechanism.
Besides, the sensitivity to $\mu(T=0)$ significantly impacts the
temperature-dependent behavior of plasmons in topological semimetals,
as evidenced by our recent research \citep{liNodallinePlasmonsZrSiX2023}.
Our study provides a convincing example of investigating the topological
properties by observing collective excitations, which is widely applicable
to other topological materials.

After clarifying the topological ground state and the resistance anomaly,
Dirac polarons could emerge as the next captivating subject in ZrTe$_{5}$.
A polaron mechanism is proposed to describe the gap opening in ZrTe$_{5}$
\citep{fuDiracPolaronsResistivity2020}, with experiments revealing
a monotonically increasing gap \citep{mohelskyTemperatureDependenceEnergy2023}.
Regarding the formation mechanism of the Dirac polarons, a Fermi liquid
system evolves into a polaronic system when its plasmon energy falls
below the optical phonon energy \citep{Verdi2017}. In our results,
the plasmon splitting distinctly demonstrates that the plasmon energy
descends below the optical phonon {[}Fig. \ref{2}(d) and (e){]} during
cooling with a non-negligible PPC, satisfying the condition for the
formation of the Dirac polarons.

\emph{Conclusion.}---We employed HREELS to study the temperature-dependent
collective excitations of ZrTe$_{5}$. Our results show that the plasmon
energy in ZrTe$_{5}$ exhibits a distinct behavior proportional to
the 1/3 power of the carrier density, which is a characteristic feature
of plasmons in 3D Dirac systems. This rules out the possibility that
ZrTe$_{5}$ is a conventional topological insulator with parabolic
bulk bands. The ground topological state of ZrTe$_{5}$ could be either
a three-dimensional Dirac semimetal (including massless or massive)
or a hyperbolic topological insulator. We attribute the origin of
the resistivity anomaly in ZrTe$_{5}$ to the inherent extrinsic-intrinsic
crossover of the Dirac systems, where the decrease in chemical potential
dominates the low-temperature extrinsic regime and thermal excitation
dominates the high-temperature intrinsic regime. We also observed
the PPC at all temperatures below 300 K, which may indicate the possible
formation of the Dirac polarons in ZrTe$_{5}$. Our study provides
a convincing example of investigating the topological properties by
observing collective excitations, which is widely applicable to other
topological materials.
\begin{acknowledgments}
This work was supported by the National Key R\&D Program of China
(No. 2021YFA1400200 and No. 2022YFA1403000), the National Natural
Science Foundation of China (No. 12274446 and No. 11874404), and the
Strategic Priority Research Program of Chinese Academy of Sciences
(No. XDB33000000). X. Z. was partially supported by the Youth Innovation
Promotion Association of Chinese Academy of Sciences. Y. S. was partially
supported by the National Natural Science Foundation of China (Grants
No. U2032204) and the Informatization Plan of Chinese Academy of Sciences(CAS-WX2021SF-0102).
\end{acknowledgments}

\bibliography{4D__Data_IOP_ZrTe5_reply_ZrTe5_manuscipt_15_ref}

\end{document}